\documentclass[onecolumn,nobibnotes,nofootinbib]{revtex4}
\usepackage{subfigure}
\usepackage{epsfig}
\usepackage{multirow}
\usepackage{color}
\usepackage{array}
\usepackage{float}
\usepackage{url}
\usepackage{amsmath,amssymb,bm}
\newcommand{\be}{\begin{equation}}
\newcommand{\ee}{\end{equation}}

\newcommand{\calN}{\mathcal{N}}
\makeatletter
\newcommand{\thickhline}{%
    \noalign {\ifnum 0=`}\fi \hrule height 1.2pt
    \futurelet \reserved@a \@xhline
}
\newcolumntype{"}{@{\hskip\tabcolsep\vrule width 1.2pt\hskip\tabcolsep}}
\makeatother
\begin{document}
\title{QCD traveling waves phenomenology revisited}
\author{J. T. Amaral ${}^1$, D. A. Fagundes${}^2$ and M. V. T. Machado ${}^3$}
\affiliation{${}^1$ Instituto de Matem\'atica, Estat\'{\i}stica e F\'{\i}sica,
Universidade Federal do Rio Grande, Caixa Postal 474
CEP 96200-970, Rio Grande, RS, Brazil
}
\affiliation{${}^2$
 Department of Exact Sciences and Education, CEE. Federal University of Santa
Catarina (UFSC) - Blumenau Campus, 89065-300, Blumenau, SC, Brazil}

\affiliation{$^3$ High Energy Physics Phenomenology Group, GFPAE. Institute of Physics,
Federal University of Rio Grande do Sul (UFRGS)
Postal Code 15051, CEP 91501-970, Porto Alegre, RS, Brazil}

\begin{abstract}
In this paper we review and update the Amaral-Gay Ducati-Betemps-Soyez saturation model, by testing it against the recent H1-ZEUS combined data on deep inelastic scattering, including heavy quarks in the dipole amplitude. We obtain that this model, which is based on traveling wave solutions of the Balitsky-Kovchegov equation and built in the momentum space framework, yields very accurate descriptions of the reduced cross section, $\sigma_{r}(x,y,Q^{2})$, as well as DIS structure functions such as $F_{2}(x,Q^{2})$ and $F_{L}(x,Q^{2})$, all measured at HERA. Additionally, it provides good descriptions of heavy quark structure functions, $F_{2}^{cc}$ and $F_{2}^{bb}$ at small-$x$ and $Q^{2}\lesssim 60$ GeV$^{2}$. We also use the
improved model to make predictions for structure functions to be measured in the near future at LHeC.

\end{abstract}
%\PACS{PACS numbers come here}
%
%%%
%
\maketitle
%

%%% Text
%
\section{Introduction}
\label{sec:intro}

Since long ago, it has been known that there is a correspondence between high energy QCD and statistical physics \cite{Munier_2009}. Of particular
importance was the discovery \cite{Munier:2003vc,Munier:2004xu,Marquet:2005ic} that at very large rapidities, $Y=\ln(1/x)$,
the leading order (fixed coupling) Balitsky-Kovchegov (BK) equation \cite{Balitsky:1995ub,Kovchegov:1999yj,Kovchegov:1999ua} is in the
universality class of the Fisher-Kolmogorov-Petrovsky-Piscounov (FKPP) equation \cite{Fisher:1937ra,Kolmogorov:1937in}, thus admitting traveling wave
solutions.
%Most of the QCD nonlinear evolution equations at very large rapidities, $Y=\ln (1/x)$, possess asymptotic solutions which fall in universality classes. Since a long time \cite{Munier:2003vc,Munier:2004xu,Marquet:2005ic}, it is known that the nonlinear Balitsky-Kovchegov (BK) equation \cite{Balitsky:1995ub,Kovchegov:1999yj,Kovchegov:1999ua} can be mapped onto the Fisher-Kolmogorov-Petrovsky-Piscounov (FKPP) equation \cite{Fisher:1937ra,Kolmogorov:1937in}.
These do not depend either on initial conditions or on the definite form of the nonlinear correction terms. Specifically, the BK equation derived for the unintegrated gluon distribution (UGD) in momentum space presents traveling-wave solutions in the transition region near the saturation domain despite the precise form of the nonlinear terms. Namely, the corresponding solution is controlled but the linear (dilute) region. Interestingly enough, the geometric scaling property observed in inclusive and exclusive processes at DESY-HERA data at small-$x$ is connected to a traveling-wave structure of the scattering amplitude for a QCD color dipole off nucleons.  The underlying quantity is the momentum saturation scale, $Q_s(Y)$,  which has its rapidity dependence driven by the velocity of the wave front, $v_c$. The evolution time is $t=\bar{\alpha}_sY$ and the position coordinate is $\rho\sim \ln (k^2/k_0^2)$ (where $k_0\sim \Lambda_{QCD}$ is a fixed infrared scale and we use $\rho$ to denote the coordinate to avoid confusion with the Bjorken $x$), and the function obeying the universal class of equations is $u(\rho,t)$. The position of the wave front is measured by the quantity  $\rho_s=\ln [Q_s^2(Y)/k_0^2]=v_cY$ and in the mean field approximation the solution to $u$ presents the form $u(\rho,t)=u(\rho-v_ct)$ \cite{Munier:2003vc,Munier:2004xu,Marquet:2005ic}.

In the large-$N_c$ limit and in the mean-field approximation, the small-$x$ behavior of the forward QCD dipole scattering amplitude, ${\cal{N}}(r, Y )$, follows the BK equation in coordinate space. This equation can be obtained also in momentum space, where it evolves
the amplitude ${\cal{N}}(k,Y)$, which is directly related to the UGD, ${\cal{F}}(k,Y)$, through
\begin{eqnarray}
\label{rel:NF}
{\cal{N}}(k,Y) = \frac{4\pi\alpha_s}{N_cR_p^2}\int_{k}^{\infty} \frac{dp}{p}{\cal{F}}(k,Y)\ln \left(\frac{p}{k}  \right),
\end{eqnarray}
where $R_p$ is the proton radius. It can be easily shown that the celebrated Golec-Biernat-Wusthoff (GBW) \cite{GolecBiernat:1999qd} form for the UGD, i.e. ${\cal{F}}_{\mathrm{GBW}}(k,Y)=F_0\,(k^2/Q_s^2)\exp[-(k^2/Q_s^2) ]$ (with $Q_s^2 = \kappa_0^2\exp(\lambda Y)$ and $F_0=N_cR_p^2/2\pi\alpha_s$) gives ${\cal{N}}_{\mathrm{GBW}}(k,Y)=\frac{1}{2}\Gamma (0,k^2/Q_s^2)$. Here, $\Gamma (0,x)$ is the incomplete Gamma function and the amplitude presents clear scaling on $\tau_s = k^2/Q_s^2(Y)$. On the other hand, the complete behavior of the amplitude at fixed QCD coupling (leading logarithmic order, LL) has been extensively investigated  and  presents a $k$-diffusion term typical of the Balitsky-Fadin-Kuraev-Lipatov (BFKL) solution \cite{Lipatov:1976zz,Kuraev:1977fs,Balitsky:1978ic} in the region $k\gg Q_s$. However, for the BK equation the saturation scale plays the role of a natural infrared cutoff and the fast broadening of the UGD at small $x$ is properly controlled. Geometric scaling behavior  on the scaling variable $\tau_s$ is restored in the region where the diffusive factor is negligible and the solution is closer to the GBW form. The BK solution at LL accuracy will be revisited in next section, where it will be used in order to describe the recent results for the proton structure function at small $x$.

Going beyond the LL approximation, the solutions of the BK equation  at next-to-leading logarithmic (NLL) order have been also investigated \cite{Peschanski:2006bm,Enberg:2006aq,Beuf:2007cw,Beuf:2007qa,Peschanski:2009ec}.  In Ref. \cite{Beuf:2007cw} three versions of the NLL BK equation were considered, namely the one-dimensional BK equation  with running coupling and two versions using quark-loop contributions. Moreover, modified BK equations including the renormalization-group corrections to the NLL BFKL kernels were studied.  It was shown that there is a unified asymptotic prediction to observables and predictions for the behavior of exact solutions fall upon a large universality class of solutions \cite{Beuf:2007cw}. That theoretical analysis led to phenomenological models presenting geometric scaling in $\sqrt{Y}$ rather than in $Y$ as in the fixed coupling case \cite{Iancu:2004es,Enberg:2005cb,Brunet:2005bz,Gelis:2006bs,Beuf:2008mf}. The role played by the fluctuations effects (Pomeron loops) in the NLL BK solution was analyzed in Ref. \cite{Beuf:2007qa}. The starting point is a Langevin equation for the forward dipole-target scattering amplitude, ${\cal{N}}(r,Y )$, with a Gaussian white noise. It was verified that a diffusive scaling for large rapidities, $Y>Y_{\mathrm{form}}$, takes place, where $Y_{\mathrm{form}}$ is the rapidity interval needed for the solution to form a wave front down to the low density domain, where the noise term is relevant. The semianalytical solution is somewhat consistent with numerical solutions of the  (1+1)-dimensional reaction-diffusion toy model for high energy QCD presented in Ref. \cite{Dumitru:2007ew}. Afterwards, this numerical solution was used to describe inclusive and diffractive deep inelastic scattering (DDIS) in \cite{Ducati:2012zi}. There, it was found that in DDIS the diffusive scaling is present for fixed coupling, and on the other hand, in the running coupling case geometric scaling takes place and it is reached at smaller values of rapidity than in the case without fluctuations \cite{Ducati:2012zi}. Furthermore, in Ref. \cite{Peschanski:2009ec} the connection between the BK equation (with nonrunning and running couplings) in the diffusive approximation with noise and the extension of the stochastic FKKP (sFKPP) to the radial wave propagation in an absorptive medium is done. An important result is that a new geometric scaling domain forward to usual traveling wave front is found. The corresponding extended scaling presents a new scaling variable, with the wave front at position ${\rho}^{\prime}=\rho-v_c\frac{t}{\sqrt[3]{\rho}}$.

In this work, we revisit the phenomenological model proposed by Amaral, Gay
Ducati, Betemps and Soyez (AGBS) \cite{deSantanaAmaral:2006fe},  based on the analytical solutions of BK equation at leading logarithmic accuracy in the momentum space. An updated AGBS model is provided through fits to the recently extracted combined HERA DIS data on the reduced cross section. Both charm and bottom quark contributions to the proton structure function, $F_2(x,Q^2)$, are included. As a by-product, charm, bottom and longitudinal structure functions are computed to be compared with the data. The plan of the paper is as follows. In Sec. \ref{sec:2} we describe the DIS cross section in terms of the AGBS model \cite{deSantanaAmaral:2006fe} for the dipole scattering amplitude in momentum space. In Sec. \ref{sec:3} fitting methods to HERA data on the reduced cross section, $\sigma_r(x,y,Q^2)$, are presented along with the fit-tuned parameters to $F^{Q\bar{Q}}$  ($Q=c,\,b$) and $F_L$ structure functions. In the last section, we discuss the main results of this study and give prospects of possible future studies. 

\section{DIS cross section in the momentum space framework}
\label{sec:2}

\subsection{DIS cross section with dipoles in momentum space}
In electron-proton DIS the $ep$ interaction is dominated by
the exchange of a virtual photon $\gamma^*$ with virtuality $Q^2$.
In the dipole model this interaction can be seen in the following way:
the virtual photon has enough energy to split into a quark-antiquark
pair, a {\it dipole}, which then interacts with the
target proton via gluon exchanges. This dipole has fixed transverse
size given by the vector $\bm{r}$, the quark carrying a fraction
$z$, and the
antiquark carrying a fraction $1-z$, of the photon longitudinal momentum. The total $\gamma^*p$ cross
section is then given by
\begin{equation}\label{eq:cross_section}
  \sigma_{T,L}^{\gamma^{*}p}(Q^2,Y)=\int
  d^2r\int_{0}^{1}dz\,\left|\Psi_{T,L}(\bm{r},z;Q^2)\right|^{2}
  \sigma_{\text{dip}}(\bm{r},Y),
\end{equation}
where $Y$ is the total rapidity interval of the $\gamma^*p$ system and $\left|\Psi_{T,L}(\bm{r},z;Q^2)\right|^{2}$ 
are the photon wave functions (well-known from
QED \cite{Nikolaev:1990ja}), which give the probabilities
for the photon, with transverse ($T$) and longitudinal ($L$)
polarization, to split into the dipole. The quantity $\sigma_{\text{dip}}(\bm{r},Y)$
is the total dipole-proton cross section which, according to the optical
theorem, is given by
\begin{equation}\label{eq:sigma-dip}
\sigma_{dip}^{\gamma^{*}p}(\bm{r},Y)=2\int d^2\bm{b}\,N(\bm{r},\bm{b},Y),
\end{equation}
where $N(\bm{r},\bm{b},Y)$ is the imaginary part of the dipole-proton scattering
amplitude in coordinate space. In the general case, the amplitude depends not only
on the dipole transverse size, but also on the impact parameter vector, $\bm{b}$, of the dipole-proton
interaction. If one neglects the $\bm{b}$ dependence (which
means considering the proton an homogeneous disk), the integration over the impact parameter is simplified. Besides, the remaining angular dependence of $\bm{r}$ can be integrated out and the dipole-proton cross section reads:
\begin{equation}\label{eq:sigmadip-r}
\sigma_{dip}^{\gamma^{*}p}(r,Y)=2\pi R_{p}^{2}N(r,Y),
\end{equation}
where $R_p$ is the proton radius and now the
amplitude depends only on the dipole size $r=|\bm{r}|$.

The above picture of DIS yields a description
of physical observables in the coordinate ($r$-dependent)
framework. An alternative approach is to express $\sigma^{\gamma^*p}$ in the momentum space framework,
where the quantities involved depend on the relative
transverse momentum of the dipole $k$. In this picture,
the dipole-target interaction is given by the dipole
scattering amplitude in momentum space, $\calN(k,Y)$, which is related to $N(r,Y)$ by the modified Fourier transform \cite{Kovchegov:1999yj,Kovchegov:1999ua}
\begin{equation}\label{eq:fourier}
 {\cal N}(k,Y)=\frac{1}{2\pi} \int \frac{d^2r}{r^2}\,e^{i\bm{k}\cdot\bm{r}}\,
 N(r,Y) = \int_0^\infty\frac{dr}{r}J_0(kr)N(r,Y).
\end{equation}
As a consequence one has, for example, that the $F_2$ proton structure function can be written as \cite{deSantanaAmaral:2006fe}:
\begin{eqnarray}\label{eq:f2-mom}
F_2(x,Q^2)&=&\frac{Q^2}{4\pi^2\alpha_{em}}
  \left[\sigma_T^{\gamma^*p}(Q^2,Y)+\sigma_L^{\gamma^*p}(Q^2,Y)\right]\nonumber\\
&=&\frac{Q^2R_p^2}{\alpha_{em}}\int_0^1 dz
\int d^2k
\left[|\tilde{\Psi}_{T}(k,z;Q^2)|^2+|\tilde{\Psi}_{L}(k,z;Q^2)|^2 \right]{\cal N}(k,Y),
\end{eqnarray}
where $\alpha_{em}$ is the electromagnetic coupling constant. $|\tilde{\Psi}_{T,L}(k,z;Q^2)|^2$ now refer to the photon wave functions in momentum space.
Their explicit forms can be straightforwardly obtained by
the relation
\begin{equation}\label{eq:wf-momsp-fourier}
    |\tilde{\Psi}_{T,L}(k,z;Q^2)|^2 = \int \frac{d^2r}{(2\pi)^2}e^{i\bm{k}\cdot\bm{r}}r^2|\tilde{\Psi}_{T,L}(r,z;Q^2)|^2
\end{equation}
and are given by \cite{deSantanaAmaral:2006fe}
\begin{eqnarray}\label{eq:wfT-momsp-explicit}
    |\tilde{\Psi}_{T}(k,z;Q^2)|^2&=&\frac{N_c\alpha_{em}}{4\pi^3}
    \sum_f e^2_q\frac{16\epsilon_q^4}{2k^2\left(k^2+4\epsilon_q^2\right)^2}
    \left\{
    [z^2+(1-z)^2]\left[\frac{4(k^2+\epsilon_q^2)}{\sqrt{k^2(k^2+\epsilon_q^2)}}
    \textrm{arcsinh}\left(\frac{k}{2\epsilon_q}\right)\right]
    \right.\nonumber\\
    &+&\left. \frac{m_f^2}{\epsilon_q^2}\left[\frac{k^2+\epsilon_q^2}{\epsilon_q^2}-
    \frac{4\epsilon_q^4+2\epsilon_q^2k^2+k^4}{\epsilon_q^2\sqrt{k^2(k^2+4\epsilon_q^2)}}
    \textrm{arcsinh}\left(\frac{k}{2\epsilon_q}\right)
    \right]
    \right\}
\end{eqnarray}
and
\begin{eqnarray}\label{eq:wfL-momsp-explicit}
    |\tilde{\Psi}_{L}(k,z;Q^2)|^2&=&\frac{N_c\alpha_{em}}{4\pi^3}
    \sum_f e^2_q\frac{16\epsilon_q^4}{2k^2\left(k^2+4\epsilon_q^2\right)^e}
     \frac{4Q^2z^2(1-z)^2}{\epsilon_q^2}\nonumber\\
     &\times&\left[\frac{k^2+\epsilon_q^2}{\epsilon_q^2}-
    \frac{4\epsilon_q^4+2\epsilon_q^2k^2+k^4}{\epsilon_q^2\sqrt{k^2(k^2+4\epsilon_q^2)}}
    \textrm{arcsinh}\left(\frac{k}{2\epsilon_q}\right)
    \right],
\end{eqnarray}
where $\epsilon_q^2=z(1-z)Q^2+m_f^2$ and $m_f$ denotes the
mass of the quark with flavor $f$.

Thus, with a model for $\calN(k,Y)$ at hand it is possible to calculate, in a momentum space framework, not only the $F_2$ structure function, but other physical quantities related to inclusive DIS, for example, the contributions
of different flavors (masses) of quarks to the $F_2$, as well as the longitudinal structure function, which can be evaluated in this momentum space approach by
\begin{equation}\label{eq:FL-momspc}
    F_L(x,Q^2) = \frac{Q^2R_p^2}{\alpha_{em}}\int_0^1 dz\int d^2k
    |\tilde{\Psi}_{L}(k,z;Q^2)|^2\calN(k,Y).
\end{equation}
The ABGS saturation model for $\calN(k,Y)$, based on the traveling wave solutions of fixed coupling BK equation, is such a model and will be reviewed below.
\subsection{Asymptotic behaviors of ${\cal N}(k,Y)$ and the AGBS model}
In the large-$N_c$ limit, the dipole scattering amplitude $\calN(k,Y)$ is the solution of BK equation
in momentum space, which can be derived from the
equation for $N(r,Y)$ by using relation (\ref{eq:fourier}). At leading order (fixed coupling) the momentum space BK equation reads (for a detailed derivation see Appendix A of
Ref. \cite{Kovchegov:1999ua})
\begin{eqnarray}
\label{eq:BK}
\partial_Y {\cal{N}} =\bar{\alpha}_s\chi(-\partial_L){\cal{N}}-\bar{\alpha}_s{\cal{N}}^{2},
\end{eqnarray}
where $\bar{\alpha}_s = \alpha_s N_c/\pi$ and  $L=\log(k^2/k_0^2)$ with $k_0$ being an infrared cutoff scale. The quantity $\chi(\gamma)=2\psi(1)-\psi(\gamma)-\psi(1-\gamma)$
is the characteristic function of the BFKL kernel \cite{Lipatov:1976zz,Kuraev:1977fs,Balitsky:1978ic}. After an appropriate change of variables,
it has been shown \cite{Munier:2003vc,Munier:2004xu} that BK equation reduces to the FKPP equation \cite{Fisher:1937ra,Kolmogorov:1937in} for $u(\rho,t)\propto {\cal{N}}(k,Y)$ when its kernel is approximated in the saddle point
approximation, i.e., to second order in the derivative $\partial_L$, the so-called diffusive approximation. In this case the equation takes the form, $\partial_{t}u(\rho,t)=\partial_{\rho}^{2}u(\rho,t)+u(\rho,t)-u^2(\rho,t)$, with $t\sim Y$ and $\rho \sim L$ corresponding to the time and space variables, respectively.

The FKKP equation presents asymptotic solutions described by traveling waves, meaning that at large times the function $u$ takes the form $u(\rho,t)=u(\rho-v_c t)$, i.e., of a front traveling to large values of $\rho$ at the speed $v_c$ without deformation. In QCD, this is
translated into the geometric scaling property,
which means that at very large rapidities (very large energies) the dipole scattering amplitude depends only on the quantity $k^2/Q_s^2$, i.e.
${\cal{N}}(k,Y)={\cal{N}}(\tau_s=k^2/Q_s^2)$.  At nonasymptotic rapidities geometric scaling is violated, and the forward amplitude takes the following form for $k\gg Q_s$ \cite{Munier:2003vc,Munier:2004xu,Marquet:2005ic},
\begin{eqnarray}
\label{eq:NNAsymp}
{\cal{N}}\left(k,Y\right) \stackrel{k\gg Q_s}{\approx}
  \left(\frac{k^2}{Q_s^2(Y)}\right)^{-\gamma_c}\log\left(\frac{k^2}{Q_s^2(Y)}\right)
\exp\left[-\frac{\log^2\left(k^2/Q_s^2(Y)\right)}{2\bar{\alpha}_s\chi''(\gamma_c)Y}\right]
\end{eqnarray}
where $\chi^{\prime\prime}$ denotes the second derivative of the BFKL kernel with respect to the anomalous dimension $\gamma$. The parameters $\gamma_c$ and $v_c$ are obtained uniquely from the BFKL kernel and
correspond to the selection of the slowest possible wave, $v_c=\bar{\alpha}_s\chi^{\prime}(\gamma_c)$. For the leading-order (LO) BFKL kernel, one obtains $\gamma_c=0.6275\ldots$, $v_c=4.88\bar{\alpha}_s$, $\chi^{\prime}(\gamma_c)=4.883\ldots$ and $\chi^{\prime \prime}(\gamma_c)=48.518\ldots$. The rapidity dependence of the saturation scale can be explicitly obtained and reads (for a up-to-date discussion see \cite{Beuf:2010aw})
\begin{equation}
\label{eq:qsat}
\log \frac{Q_s^2(Y)}{k_0^2}=
v_cY-\frac{3}{2\gamma_c}\log Y
-\frac{3}{\gamma_c^2}\sqrt{\frac{2\pi}{\bar{\alpha}_s\chi^{\prime\prime}(\gamma_c)}}\frac{1}{\sqrt{Y}}\mathcal{O}(1/Y).
\end{equation}
From Eq. (\ref{eq:NNAsymp}) it is possible to verity that
the  geometric scaling is obtained for a kinematic range where $k^2\lesssim Q_s^2(Y)e^{\beta \sqrt{Y}}$ (the so-called geometric scaling window), with $\beta = \sqrt{2\chi^{\prime \prime}(\gamma_c)\bar{\alpha}_s}$.

The results described above motivated the construction
of the AGBS saturation model \cite{deSantanaAmaral:2006fe}, which explores the implications of the traveling wave solutions of BK evolution equation to the $\gamma^*p$ scattering. It provides the following phenomenological expression for ${\cal{N}}(k,Y)$:

%a phenomenological model for the amplitude
%$\calN(k,Y)$ based on the traveling wave solutions of BK
%equation, the AGBS model \cite{deSantanaAmaral:2006fe}.
%The expression proposed for the amplitude is}
%The AGBS model \cite{deSantanaAmaral:2006fe} is an approach which explores the implications of the traveling wave solutions of BK evolution equation to the $\gamma^*p$ scattering. It provides a phenomenological expression for ${\cal{N}}(k,Y)$ which analytically interpolates between the behaviors of the amplitude at the saturated and dilute domains, Eqs.(\ref{eq:NNAsymp}) and (\ref{eq:Nsat}). The final expression for ${\cal{N}}(k,Y)$ \cite{deSantanaAmaral:2006fe} reads
 \begin{equation}\label{eq:Tmodel}
{\cal N}^{\rm{AGBS}}(k,Y) = \left[\log\left(\frac{k}{Q_s}+\frac{Q_s}{k}\right)+1\right]\left(1-e^{-T_{\text{dil}}}\right),
\end{equation}
where
\begin{equation}\label{eq:Tdil}
T_{\text{dil}} = \exp\left[-\gamma_c\log\left(\frac{k^2}{Q_s^2(Y)}\right)
-\frac{{\cal L}^2-\log^2(2)}{2\bar{\alpha}_s\chi''(\gamma_c)Y}\right],
\end{equation}
\begin{equation}\label{eq:l_red}
{\cal L}=
\log \left[1+\frac{k^2}{Q_s^2(Y)}\right]
\end{equation}
and the saturation scale is given by
only by the dominant term of
 Eq.(\ref{eq:qsat}), i.e.,
 \begin{equation}\label{eq:qsat-approx}
     Q_s(Y)\approx k_0^2e^{v_cY}.
 \end{equation}
The AGBS model, given by Eqs. (\ref{eq:Tmodel}) - (\ref{eq:qsat-approx}), provides an analytical interpolation between the dilute region, given
by Eq.(\ref{eq:NNAsymp}), where $k\gg Q_s$ (the tail
of the wave front), the region around the saturation
scale, where $k\approx Q_s$, and the deep saturation region, where 
$k\ll Q_s$. In a rough approximation, the expression
for the amplitude in the saturation region has been proposed to be the Fourier transform (\ref{eq:fourier}) of a Heaviside function  ${\cal{N}}(r,Y)=\Theta (rQ_s-1)$, which yields the following behavior
of the amplitude in momentum space, in the region $k\ll Q_s$ \cite{deSantanaAmaral:2006fe}:
\begin{eqnarray}
\label{eq:Nsat}
{\cal{N}}\left(\frac{k}{Q_s(Y)},Y\right)\stackrel{k\ll Q_s}{=} a - \log\left(\frac{k}{Q_s(Y)}\right)
\end{eqnarray}
with $a$ being a constant to be determined by the boundary conditions \footnote{This behavior can be also obtained from the explicit solution of BK evolution equation inside the saturation region. Indeed, in the region $Q_s\gtrsim k\gg \Lambda_{\mathrm{QCD}}$, a similar expression for $\calN(k,Y)$ can be derived from the Levin-Tuchin (LT) formula \cite{Levin:1999mw,Levin:2000mv} for the S matrix valid for larger dipoles,  $r \gtrsim 1/Q_s$. Starting from the LT solution,
\begin{eqnarray}
S(r,Y) = \exp \left(-\tau \ln^2[r^2Q_s^2] \right),
\end{eqnarray}
the corresponding UGD has been recently obtained in Ref. \cite{Abir:2018hvk}. In the leading logarithmic  approximation and for $k^2 \ll Q_s^2$ it can be approximated to \cite{Abir:2018hvk}
 \begin{equation}
 {\cal{F}}(k,Y) 
  \approx  \frac{N_cR_p^2 \tau}{\pi\alpha_s} \ln \left(\frac{k^2}{4Q_s^2}  \right) \exp \left[-\tau \ln^2\left(\frac{k^2}{4Q_s^2}\right)   \right],
 \end{equation}
where $\tau$ is a constant well determined. Using the relation (\ref{rel:NF}), it is straightforward to
show that the dipole scattering amplitude in momentum space is given by
 \begin{eqnarray}
 {\cal{N}}(k,Y) & \stackrel{Q_s\,\gtrsim k\,\gg \Lambda_{\mathrm{QCD}}}{\approx} & \frac{1}{4\sqrt{\tau}}\left[  \sqrt{\pi}\,\mathrm{erf}\left( \sqrt{\tau}\ln \frac{k^2}{4Q_s^2} \right) + 2\sqrt{\tau}\ln \left(\frac{4Q_s^2}{k^2}\right)\right],\\
& \stackrel{k\ll Q_s}{\approx} & \frac{1}{4\sqrt{\tau}}\left[  -\sqrt{\pi} + 4\sqrt{\tau}\ln \left(\frac{2Q_s}{k}\right)\right],
 \end{eqnarray}
 which has exactly the same parametric form as the simple asymptotic expression in Eq. (\ref{eq:Nsat}).}. An important comment
 is in order: as explained in \cite{deSantanaAmaral:2006fe}, the eikonal
 way of unitarization of $\calN^{\textrm{AGBS}}$,
 $1-e^{-T_{\textrm{dil}}}$, is not physically motivated,
 it has been chosen because of its simplicity
 [for example, the form $T_{\textrm{dil}}/(1+T_{\textrm{dil}})$
 would work equally well].

In the paper where the model was proposed \cite{deSantanaAmaral:2006fe}, the AGBS model was used to fit measurements of the $F_2$ proton structure function from H1 \cite{Adloff:2000qk} and ZEUS Collaborations \cite{Breitweg:2000mu,Chekanov:2001qu}
taking heavy-quark (charm) effects into account. Afterwards, in \cite{Basso:2011fb} another
fit to $F_2$ has been performed, considering only the contribution of light quarks, but using (more
recent) H1 and ZEUS combined HERA data \cite{Abramowicz:2015mha}.  The model has also been also used to investigate possible pomeron loop effects at HERA \cite{Basso:2008re} and to describe
inclusive hadron and photon production at the LHC \cite{Basso:2012nb}. Thus, besides being useful in the description of DIS data, it also provides the fundamental tools to study inclusive observables at RHIC and LHC energies.  In all these
phenomenological applications the AGBS model has been shown to be successful in the description of the data . This, together with the fundamental properties underlying the construction of the model, makes its improvement an interesting issue. This will be done
in what follows.

\section{DIS data and fitting procedure}
\label{sec:3}
%\subsection{Data set and parameters}

In this paper, we make an improvement of the AGBS model by updating its parameters with a fitting
procedure to recent high-precision HERA data \cite{Abramowicz:2015mha}, including heavy -- charm and bottom -- quarks. In particular, we fit the reduced cross section data \cite{Abramowicz:2015mha}, which reads
\begin{equation}\label{eq:red_xsec}
\sigma_{r}(x,y,Q^{2})=F_{2}(x,Q^{2})-\frac{y^{2}}{1+(1-y)^2}F_{L}(x,Q^{2}).
\end{equation}
where $y = Q^2/(sx)$ is the inelasticity variable, $\sqrt{s}$ denotes the center of mass energy of the $ep$ collision and $F_{L}(x,Q^{2})$ is the longitudinal structure function.

In fitting $\sigma_{r}$ a kinematic cut to HERA data is applied to the Bjorken-$x$ variable, namely $x\leq 0.01$, since this approach is
conceived to describe high-energy amplitudes (the small-$x$ behavior). Two bins of the photon virtuality
are considered:
\begin{equation}\label{eq:Q2bins}
\left\{\begin{array}{cc}
  Q^2 \in [0.045,45] \textrm{ GeV}^2\ (\text{bin 1} ) & \textrm{and}  \\
                       &                 \\
  Q^2 \in [0.045,150] \textrm{ GeV}^2\ (\text{bin 2}). &
\end{array}\right.
\end{equation}
Both bins prevent us from the need to include
Dokshitzer-Gribov-Lipatov-Altarelli-Parisi (DGLAP) corrections, which must be properly accounted for at too high values
of $Q^2$. The choice of fitting data in the bin 1 range can be regarded as a conservative one, with respect to traditional approaches such as, e.g., GBW \cite{Golec-Biernat:2017lfv}, for which an even lower $Q^{2}_{max}$ (=10 GeV$^{2}$, as long as DGLAP corrections are not included) is probed. Moreover, as we take into account heavy quark contributions and since the experimental range considered includes very small values of $Q^2$, we perform the usual kinematic shift in the
definition of Bjorken-$x$ \cite{GolecBiernat:1999qd}
\begin{equation}\label{eq:modif-x}
    x\to \tilde{x}_{f}=x\left(1+\frac{4m_f^2}{Q^2}\right).
\end{equation}
for charm and bottom, when the cut $\tilde{x}_{f}\leqslant 0.1$ is satisfied. Otherwise the contribution of heavy quarks is switched off. 

Fits have been performed using the ROOT framework \cite{Brun:1997pa,Antcheva:2011zz}, through the members of the TMINUIT class \footnote{URL: \url{https://root.cern.ch/doc/master/classTMinuit.html}}. In specific, we use the MIGRAD algorithm throughout, setting the confidence level (CL) to 95$\%$ \footnote{As it is widely known the UP parameter in MINUIT  may vary according to the number of degrees of freedom and the confidence level. In our case, with four or five fit parameters one uses, 9.49 and 11.07, respectively.}.  Goodness-of-fit is evaluated using the standard chi-squared  ($\chi^{2}$) per degrees of freedom ($\text{d.o.f.}$) criterion, with
\begin{equation}
\chi^{2}=\sum_{i=1}^{N_{p}} \frac{1}{\sigma_{i}^{2}}(s_{i}(x_{i},y_{i},Q_{i}^{2})-\sigma_{r}(x,y,Q^{2}))^{2},    
\end{equation}
$s_{i}$ representing the reduced cross section data ($N_{p}=524$ for bin 1 and $N_{p}=659$ for bin 2), $\sigma_{i}$ the total uncertainty with respect to central values, $s_{i}$, and $\sigma_{r}(x,y,Q^{2})$ our model, according to Eqs. (\ref{eq:f2-mom}), (\ref{eq:FL-momspc}) and (\ref{eq:red_xsec}). We also provide the integrated probability, $P(\chi^{2}; \text{d.o.f.})$, the well-known $p-$value, also as goodness-of -fit estimator, with due care, namely limiting to interpret its results in the light of an overall agreement with data sets for the various models tested, specially when comparing fits to bin 1 and bin 2, and not in the traditional sense, that is, as a test of hypothesis used to discriminate good from bad models.

Concerning the model parameters, the one kept fixed in this analysis is $\bar{\alpha}_s=0.2$.  For the value of the critical slope $\gamma_c$, two
different scenarios were tested: $\gamma_c=0.6275$, which as mentioned before comes from
the LO BFKL kernel, and $\gamma_c$ considered as a free parameter, a case which was tested in the fit
performed in \cite{Soyez:2007kg} using the Iancu-Itakura Munier (IIM) saturation model for $N(r,Y)$ including the heavy quarks. The value obtained in \cite{Soyez:2007kg} was $\gamma_c=0.7376$,
in agreement with what is expected from NLO BFKL ($\gamma_c \gtrsim 0.7$).
Thus, as in the previous studies using AGBS model, we are left with at least four
free parameters, $v_c$, $k_0^2$, $R_p$ and $\chi^{\prime\prime}(\gamma_c)$.
For the rapidity dependence of the saturation
scale $Q_s$ we keep only the first (leading) term, see Eq.(\ref{eq:qsat-approx}), just
as it was done in the original work  and all other studies which used AGBS model for data description. Clearly, keeping only the leading term is a phenomenological choice, since the amplitude (\ref{eq:Tmodel}) is not a solution to LO BK equation, but a model based on the behavior of its solutions in asymptotic regimes.

\begin{table}[ht]
\centering
\caption{Parameters obtained from the fits in Refs.\cite{deSantanaAmaral:2006fe,Basso:2011fb}.
Only the results which provided the best fit quality are presented.}
\begin{tabular}{c|c|c|c|c|c|c|c|c}
\thickhline
Work  & $m_q$ [GeV] & $m_c$ [GeV] & $m_b$ [GeV] & $k_0^2$ ($10^{-3}$ GeV$^2$) & $v_c$ & $\chi^{\prime\prime}(\gamma_c)$ & $R_p$ (GeV$^{-1}$)  & Fit quality\footnote{$\chi^2/\textrm{n.o.p.}$=$\chi^2$ per number of points.} \\
\thickhline
Ref.\cite{deSantanaAmaral:2006fe}& 0.05 & 1.3 & $\cdots$
& 7.155 $\pm$ 0.624 &  0.193 $\pm$ 0.003 &2.196 $\pm$ 0.161  &   3.215 $\pm$ 0.065 & $\chi^2/\textrm{n.o.p.} =0.988$  \\
\hline
Ref.\cite{Basso:2011fb}  & $0.14$ & $\cdots$ & $\cdots$
                                  & 1.13 $\pm$ 0.024
                            
                                  & 0.165 $\pm$ 0.002      & 7.488 $\pm$ 0.081
                                  & 5.490  $\pm$ 0.039 & $\chi^2/\textrm{d.o.f.} =0.903$\\
\thickhline
\end{tabular}
\label{tab:old-res}
\end{table}

 For the quark masses we consider two different situations: (\texttt{i}) with only light quarks and (\texttt{ii}) with light and heavy (charm and bottom) quarks. In both situations we use two different values for the light quark masses: $m_q=m_{u,d,s}=0.14$ and $0.05$ GeV. The first value is the most used in DIS phenomenology in the dipole framework, while the second
is the one which provided the best fit to previous (not
combined) HERA data in the original AGBS
model \cite{deSantanaAmaral:2006fe}. In the case
where heavy quarks are taken into account, charm
and bottom quark masses are assumed to be $m_c=1.3$ GeV and $m_b=4.6$ GeV, respectively.
For the sake of comparison, we show the values obtained 
in Refs. \cite{deSantanaAmaral:2006fe} (light quarks and charm quarks) and \cite{Basso:2011fb} (only light quarks)
in Table \ref{tab:old-res} (we present only the main results).

The main results described above are presented in Tables \ref{tab:params_fits_light} and \ref{t:params_fits}, where we introduce
labels for different fit variants: $V_{i}B_{j}$, with $i=1, 2, 3, 4$ standing for the different values of light quark masses in both situations described above, while $j=1,2$ indicates which bin has been used to tune the model parameters.

%%%%%%%%%

\begin{table}[ht]
\centering
\caption{Parameters obtained from the fits performed
using only light quark masses. Only the results
with best $\chi^2/\textrm{d.o.f.}$ are shown. Fit variants are indicated by $V_{i}B_{j}$, with $i=1, 2$ standing for different values of light quark masses, $m_{q}$ (fixed), while $j=1, 2$ indicates which bin have been used to tune parameters of each model.}
%\vspace*{.2cm}
\resizebox{\textwidth}{!}{

\begin{tabular}{c|c|c|c|c|c|c|c|c}
\thickhline
Bin & Variant & $m_q$ {[}GeV{]} & $k_0^2$ ($10^{-3}$ GeV$^2$)& $v_c$  & $\chi^{\prime\prime}(\gamma_c)$ &  $R_p$ (GeV$^{-1}$) & $\chi^2/$dof & $p-$value \\
\thickhline
\multirow{2}{*}{{$Q^{2}:\ $[}0.045,150{]} GeV$^{2}$}
 &  $V_1B_2$ & $0.14$  & $ 1.19  \pm 0.57$ & $ 0.171  \pm 0.012 $ & $ 7.4 \pm 1.3$ & $ 5.32 \pm 0.74 $           & 608.269/655 =  0.929&  0.904    \\
\cline{2-9}
 & $V_2B_2$ & 0.05 &  $1.99 \pm 0.74$ & $ 0.1978\pm 0.0093$ &  $6.4  \pm 1.0$ & $4.26 \pm 0.42$  &  606.922/655 = 0.927 &  0.910    \\
\thickhline
\end{tabular}
\label{tab:params_fits_light}
}
\end{table}

In Table \ref{tab:params_fits_light} we show the best results (Bin 2) of
our fits to the DESY-HERA data for the reduced cross section when only light quarks considered.
Although the two
different choices of quark masses lead to significant
differences with respect to the resulting values of the parameters of the AGBS model, they provide fits with similar qualities. In order to perform a cross-check, we can compare our results
with those obtained from the fits
of \cite{Basso:2011fb} (see Table \ref{tab:old-res}). We see that the results are quite similar (concerning both the parameter values and the quality of the fit) to those obtained in the present work, with the same value, $m_q=0.14$ GeV, for the light quark masses.

\begin{table}[ht]
\centering
\caption{Best fit parameters of our model in the bins 1 and 2 taking into account all quark flavors. As before, fit variants are indicated by $V_{i}B_{j}$, with $i=3, 4$ standing for different values of light quark masses. The four free parameters are shown for the fit variants, along with their dimensions. Uncertainties are given within $95\%$ confidence level, along with $\chi^{2}$/d.o.f. and $p-$value in each variant. In all cases presented
the parameter $\gamma_c$ is kept fixed at the value $\gamma_c = 0.6275$.}
%\vspace*{.2cm}
\resizebox{\textwidth}{!}{
\begin{tabular}{c|c|c|c|c|c|c|c|c|c|c}
\thickhline
Bin &  Variant & $m_{q}$ {[}GeV{]} & $m_{c}$ {[}GeV{]} & $m_{b}$ {[}GeV{]} & $k_{0}^{2}$ ($\times 10^{-3}$) {[}GeV$^{2}${]} & $v_{c}$  & $\chi''_{c}$ & $R_{p}$ {[}GeV$^{-1}${]} & $\chi^{2}$/dof & $p-$value \\ \thickhline
%%%%%%%%%%%%%%%%%%%%%%%%%%%%%%%%%%%%%%%%%%%%%%%%%%%%%%%%%%%%%%%%%%%%%%%%%%%%%%%%%%%%%%%%%%%%%%%%%%%%%%%
%bin1
\multirow{2}{*}{{$Q^{2}:\ $[}0.045,45{]} GeV$^{2}$}
%&  $V_{1}B_{1}$ & 0.14  & $-$ & $-$  &  $ 1.33 \pm 0.95$ & $ 0.166 \pm 0.015$ & $ 7.0 \pm 2.0 $ & $5.3\pm  1.0 $           & 204.944/267 = 0.768   & 0.998 \\  \cline{2-11} 
 & $V_{3}B_{1}$ & 0.14  & 1.3 & 4.6  & $2.6 \pm 2.0$ &  $0.136  \pm  0.024$ & $3.40 \pm 1.3$ & $ 5.1 \pm  1.2$  & 459.429/520 =  0.884 & 0.974\\ \cline{2-11}

%& $V_{3}B_{1}$ & 0.05 & $-$ & $-$ &  $ 2.073 \pm 0.169$ & $ 0.1944 \pm 0.0062 $ & $ 6.21 \pm 0.33$ & $ 4.270 \pm 0.054$  &  207.976/267 = 0.779 &  0.997\\ \cline{2-11} 

&  $V_{4}B_{1}$ & 0.05  & 1.3 & 4.6  & $ 4.12 \pm 0.59$ &  $0.1619 \pm 0.0071$  & $2.91 \pm 0.26$ & $4.04 \pm  0.18$ & 453.791/520 = 0.873 &  0.983 \\
 \thickhline
%%%%%%%%%%%%%%%%%%%%%%%%%%%%%%%%%%%%%%%%%%%%%%%%%%%%%%%%%%%%%%%%%%%%%%%%%%%%%%%%%%%%%%%%%%%%%%%%%%%%%%%
%bin2
\multirow{2}{*}{{$Q^{2}:\ $[}0.045,150{]} GeV$^{2}$}
%&  $V_{1}B_{2}$ & 0.14  & $-$ & $-$  & $ 1.331  \pm 0.091$ & $ 0.1660  \pm 0.0056$ & $ 7.00 \pm 0.26$ & $ 5.269 \pm 0.076 $           & 265.64/305 =  0.871 &  0.950 \\  \cline{2-11} 
  & $V_{3}B_{2}$ & 0.14  & 1.3 & 4.6  & $ 1.704 \pm  0.075$ & $0.1380 \pm 0.0035$  & $ 4.04 \pm 0.13$ & $5.603 \pm  0.086$  & 818.47/655 =  1.25 &  $1.30\times 10^{-5}$ \\ \cline{2-11}
%  & $V_{3}B_{2}$ & 0.05 & $-$ & $-$ &  $2.26 \pm 0.59$ & $ 0.1947\pm 0.0098$ & $5.965  \pm 0.71$ & $4.19 \pm 0.30$  &  266.30/305 = 0.873 &  $0.946$ \\ \cline{2-11} 
 & $V_{4}B_{2}$ & 0.05  & 1.3 & 4.6  & $2.88 \pm 0.14$  & $0.1697  \pm 0.0059$ & $3.4378 \pm 0.0054$ & $4.36 \pm 0.13$  & 781.986/655 = 1.19  & $4.46\times 10^{-4}$   \\
 \thickhline
%%%%%%%%%%%%%%%%%%%%%%%%%%%%%%%%%%%%%%%%%%%%%%%%%%%%%%%%%%%%%%%%%%%%%%%%%%%%%%%%%%%%%%%%%%%%%%%%%%%%%%%
\end{tabular}
\label{t:params_fits}
}
\end{table}

In Table \ref{t:params_fits} we summarize the results of the fits to the data on the reduced cross section data with light and heavy quarks for the two $Q^2$ bins.
A suited fit quality is found (see variants $V_3B_2$ and $V_4B_2$), given the data precision and a minimal number of fitted parameters. In the Table \ref{t:params_fits}, we present only the results with the parameter $\gamma_c$ fixed at the value $\gamma_c=0.6275$, since it provided the best fits to the data. As mentioned before, we have tested the case where $\gamma_c$ is left free and verified a good stability for this parameter, with the fitted one being very close to the that coming from BFKL dynamics.
The updated parameters are close to the original ones (Ref. \cite{deSantanaAmaral:2006fe}, with only charm effects taken into account, see Table \ref{tab:old-res}) with $v_c$ having lower values by around 15$\%$.
We clearly see that the inclusion of heavy quarks still provide good fits to HERA data (see Tables \ref{tab:params_fits_light} and \ref{t:params_fits}). The bottom quark contribution plays a small role in the bin 1, whereas in the bin 2 it is significant, although most of the results present $p$-values larger than the confidence level considered, $\alpha = 0.05$,  which demonstrate good statistical significance of the analysis. The variability in the fit quality estimators, $\chi^{2}/\textrm{d.o.f.}$ and $p-$ values, between the fits performed in bins 1 and 2 can be noticeable, even though that does not compromise the goodness of fits by all means. In fact, as we shall see in the following, fits and predictions of models obtained by tuning our model parameters with bin 1 essentially overlap with the ones from bin 2. Such behavior, seems to evidence not only an important effect of high$-Q^{2}$ and high$-x$ in our dipole amplitude, but also that fitting a larger $Q^{2}$ bin may not be required in order to obtain reasonable predictions for heavy quarks structure functions. For that reason, despite the low $p-$values shown in Table III for bin 2 fits, we keep those results with $\chi^{2}/\textrm{d.o.f.}\sim 1$, as they still provide good fits (both visually and statistically) for such a large number of degrees of freedom ($\sim 650$), since this worsening with respect to bin 1 fits can be traced to the effect of including heavy quarks in the amplitude in a larger $Q^{2}$ range.

In essence, these results demonstrate that the AGBS model remains doing a good job even at large virtualities and small-$x$, mimicking part of the typical DGLAP evolution (driven by the extended geometric scaling behavior present in large $k$ tale of the dipole amplitude). Parameter $v_c=\lambda \simeq 0.15-0.17$ is compatible with $\lambda$ values found in recent analyses using dipole models with extended geometric scaling in coordinate space. For instance, IIM/CGC model \cite{Soyez:2007kg,Rezaeian:2013tka} gives $\lambda \simeq 0.23$ whereas b-CGC model  \cite{Rezaeian:2013tka} found $\lambda = 0.2063$. The value of the parameter $R_p\simeq 4.62-5.3$ GeV$^{-1}$, which is  related to the black disc limit of $\gamma^*p$ cross section, $\sigma_0=2\pi R_p^2\simeq 52-67$ mb,  produces  larger values compared to corresponding models in coordinate space where $\sigma_0\sim 30$ mb \cite{Rezaeian:2013tka,Luszczak:2016bxd,Golec-Biernat:2017lfv}.  

\begin{figure}[H]
    \centering
    \includegraphics[scale=0.8]{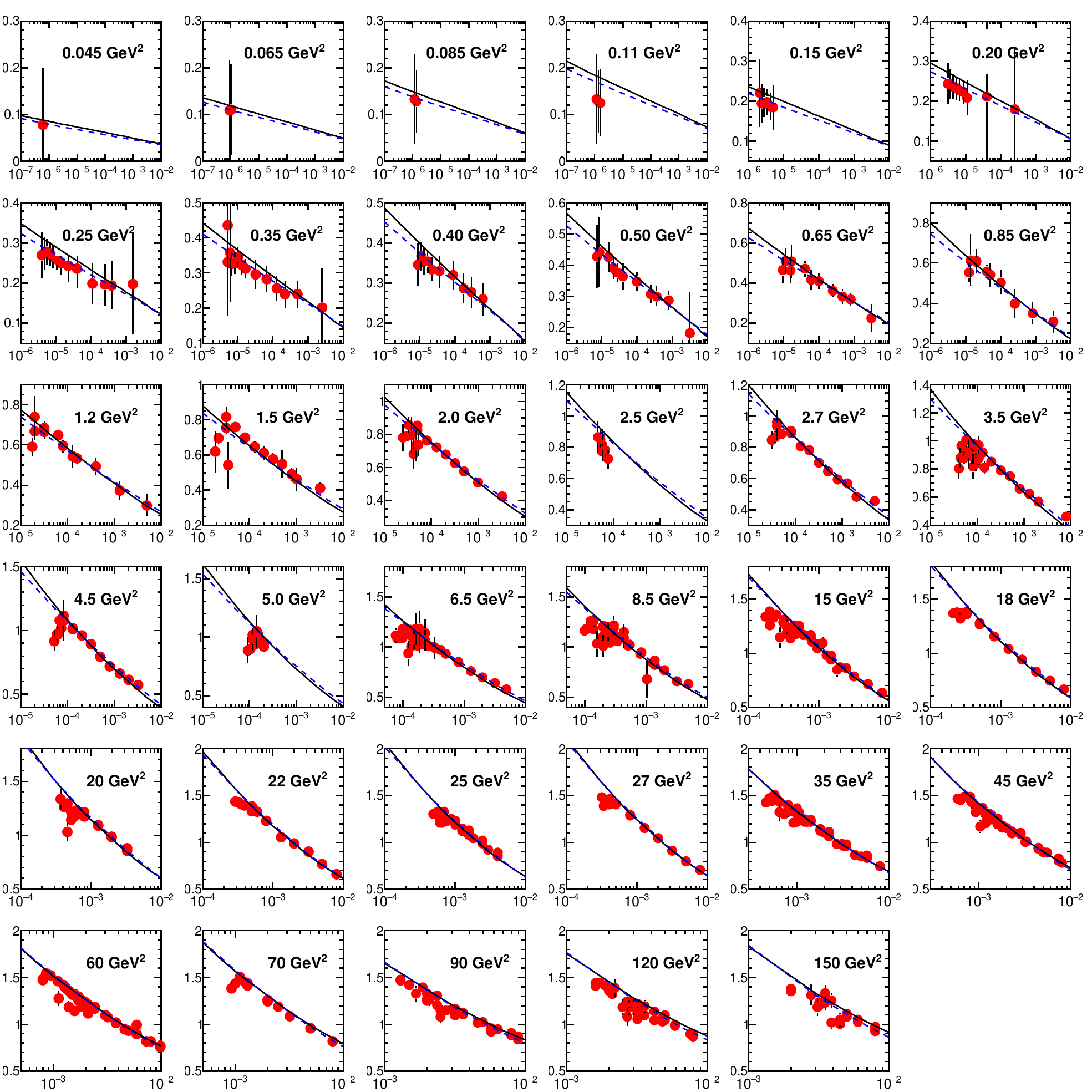}
    \caption{ \textit{Red circles}: H1-ZEUS $e^{\pm}p$ combined $F_{2}(x,Q^{2})$ data  in the range $x \leqslant 0.01$ and 0.045 GeV$^{2}$ $\leqslant Q^{2} \leqslant 150$ GeV$^{2}$ \cite{Abramowicz:2015mha}. $F_{2}$ uncertainties are estimated, considering $\delta F_{2} \approx \delta \sigma_{r}$. \textit{Curves}: black solid and blue dashed curves are the predictions of variants  $V_{1}B_{2}$ and $V_{3}B_{2}$, following from fits to $\sigma_{r}(x,y,Q^{2})$ including heavy quarks and only with light ones. Fit parameters of these curves are given in Table \ref{t:params_fits}.}
    \label{fig:F2}
\end{figure}

In Fig. \ref{fig:F2}, a comparison between the variants $V_1B_2$ (solid lines, only light quarks)  and $V_3B_2$ (dashed lines, including charm and bottom) is shown against the H1-ZEUS combined $F_2$ data at $Q^2\in [0.1,150]$ GeV$^2$ and $x\leq 10^{-2}$. A very good agreement with data can be observed and the curves are practically the same at very low-$Q^2$. Small deviations appear only at large $Q^2$ and very small $x$. The results for light quarks are steeper ($v_c\simeq 0.17$)  than for those including heavy quarks ($v_c\simeq 0.15$). The resulting dipole amplitude in momentum space obtained from present fits can be used for the prediction of LHC cross sections along the lines presented in Ref. \cite{Basso:2011fb}. In addition, by using of Eq.(\ref{rel:NF}), the proton unintegrated gluon distribution can be easily obtained. This is important for the physics based on calculations in the scope of TMD/$k_{\perp}$-factorization formalism. 

As previously stated, the ABGS model nicely describes HERA data for small and moderate photon virtualities including the transition of the DIS structure functions to small values of $Q^2$. It is known that this is achieved by the parton saturation corrections to the BFKL formalism embedded in the approach. This should be more evident in the longitudinal structure function, which is strongly affected by the screening corrections.

\begin{figure}[H]
    \centering
    \includegraphics[scale=0.7]{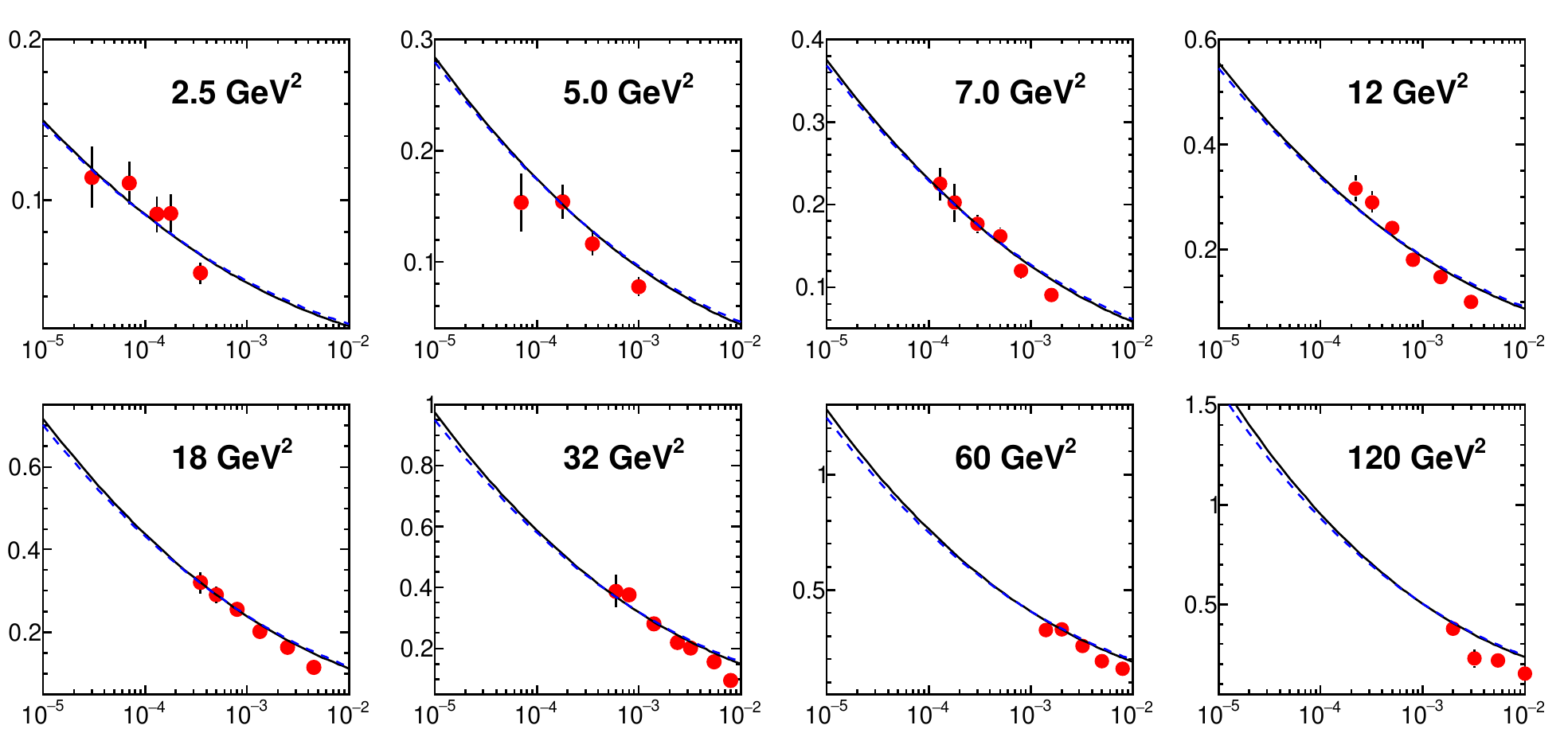}
    \caption{ Charm structure function, $F_{2}^{cc}(x,Q^{2})$, estimates from HERA in the range 2.5 GeV$^{2}$ $\leqslant Q^{2} \leqslant 120$ GeV$^{2}$ \cite{H1:2018flt}, assuming $F_{2}^{cc}\approx\    \sigma_{r}^{cc}$.  Predictions of variants  $V_{3}B_{1}$ and $V_{3}B_{2}$  are given by  blue dashed and black solid  curves, respectively.}
    \label{fig:F2cc}
\end{figure}

\begin{figure}[H]
    \centering
    \includegraphics[scale=0.7]{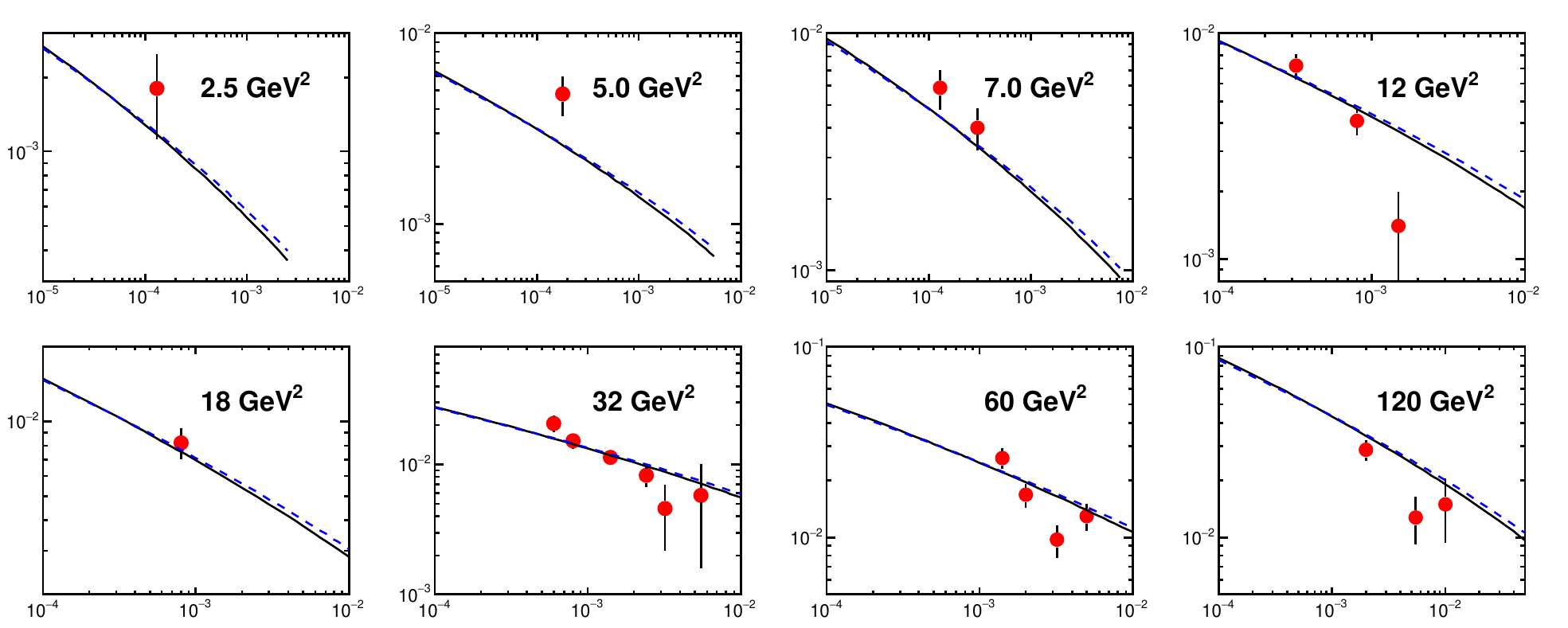}
    \caption{ Bottom structure function, $F_{2}^{bb}(x,Q^{2})$, estimates from HERA in the range 2.5 GeV$^{2}$ $\leqslant Q^{2} \leqslant 120$ GeV$^{2}$ \cite{H1:2018flt}, assuming $F_{2}^{bb}\approx\    \sigma_{r}^{bb}$. Predictions of variants $V_{3}B_{1}$ and $V_{3}B_{2}$  are given by  blue dashed and black solid  curves, respectively.}
    \label{fig:F2bb}
\end{figure}

\begin{figure}[H]
    \centering
    \includegraphics[scale=0.7]{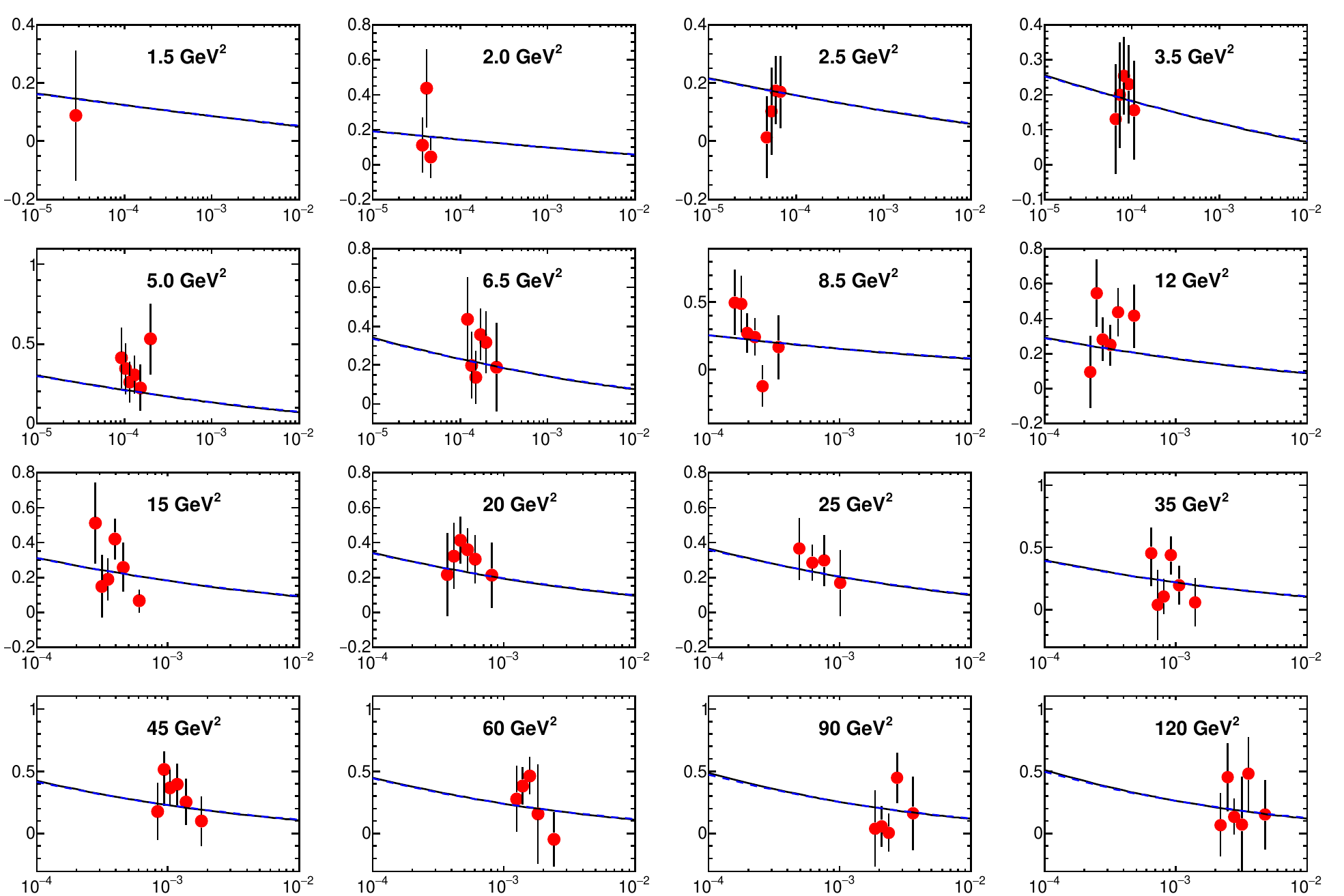}
    \caption{ Longitudinal structure function, $F_{L}(x,Q^{2})$,  from HERA in the range 1.5 GeV$^{2}$ $\leqslant Q^{2} \leqslant 120$ GeV$^{2}$ \cite{Andreev:2013vha} Predictions of  variants $V_{3}B_{1}$ and $V_{3}B_{2}$  are given by  blue dashed and black solid  curves, respectively.}
    \label{fig:FL}
\end{figure}

With the parameters given in Table \ref{t:params_fits}, we are
able to compute and make predictions for the charm and bottom
structure functions. The results are presented in Figs. \ref{fig:F2cc} and \ref{fig:F2bb}, respectively, where both contributions for the $F_2$ structure function are considered in the range 2.5 GeV$^{2}$ $\leqslant Q^{2} \leqslant 120$ GeV$^{2}$ \cite{H1:2018flt}, and we use the variants  $V_3B_1$
and $V_3B_2$. In the case of charm structure
function, $F_2^{c\bar{c}}$, one  sees that AGBS model provides
a good description of the data within a wide range of the
photon virtuality, up to 60 GeV$^2$, and a reasonable
description at $Q^2=120$ GeV$^2$.  We have also made predictions for the bottom structure function, $F_2^{b\bar{b}}$, finding a good
agreement with the data, in particular for $Q^2\geqslant 7$
GeV$^2$, where there is a larger number of experimental points.
Finally, we present our predictions for the longitudinal structure function, $F_L(x,Q^2)$, which in the present analysis can be
evaluated using Eqs. (\ref{eq:wfL-momsp-explicit}) and (\ref{eq:FL-momspc}). The results are presented in Fig. \ref{fig:FL}, where we show the behavior of $F_L$ as a function of $x$ in the range 1.5 GeV$^{2}$ $\leqslant Q^{2} \leqslant 120$ GeV$^{2}$ of the photon virtuality, considering, as before, variants
$V_{3}B_{1}$ and $V_{3}B_{2}$. In all the ranges considered, we see that
AGBS model provides a good agreement with the data. Besides, as already mentioned, Figs. \ref{fig:F2cc} and \ref{fig:F2bb} reveal that bin 1 data are sufficient to furnish accurate descriptions of both,  $F_2^{c\bar{c}}$ and $F_2^{b\bar{b}}$, even at virtualities as large as $Q^{2}\sim 100$ GeV$^{2}$. Notwithstanding, this effect is even more drastic for $F_L(x,Q^2)$, as one can see from Fig. \ref{fig:FL}, in which $V_{3}B_{1}$ and $V_{3}B_{2}$ exactly overlap.

%\subsection{Prediction for charm, bottom and longitudinal structure functions}

\begin{figure}[H]
    \centering
    \includegraphics*[scale=0.45]{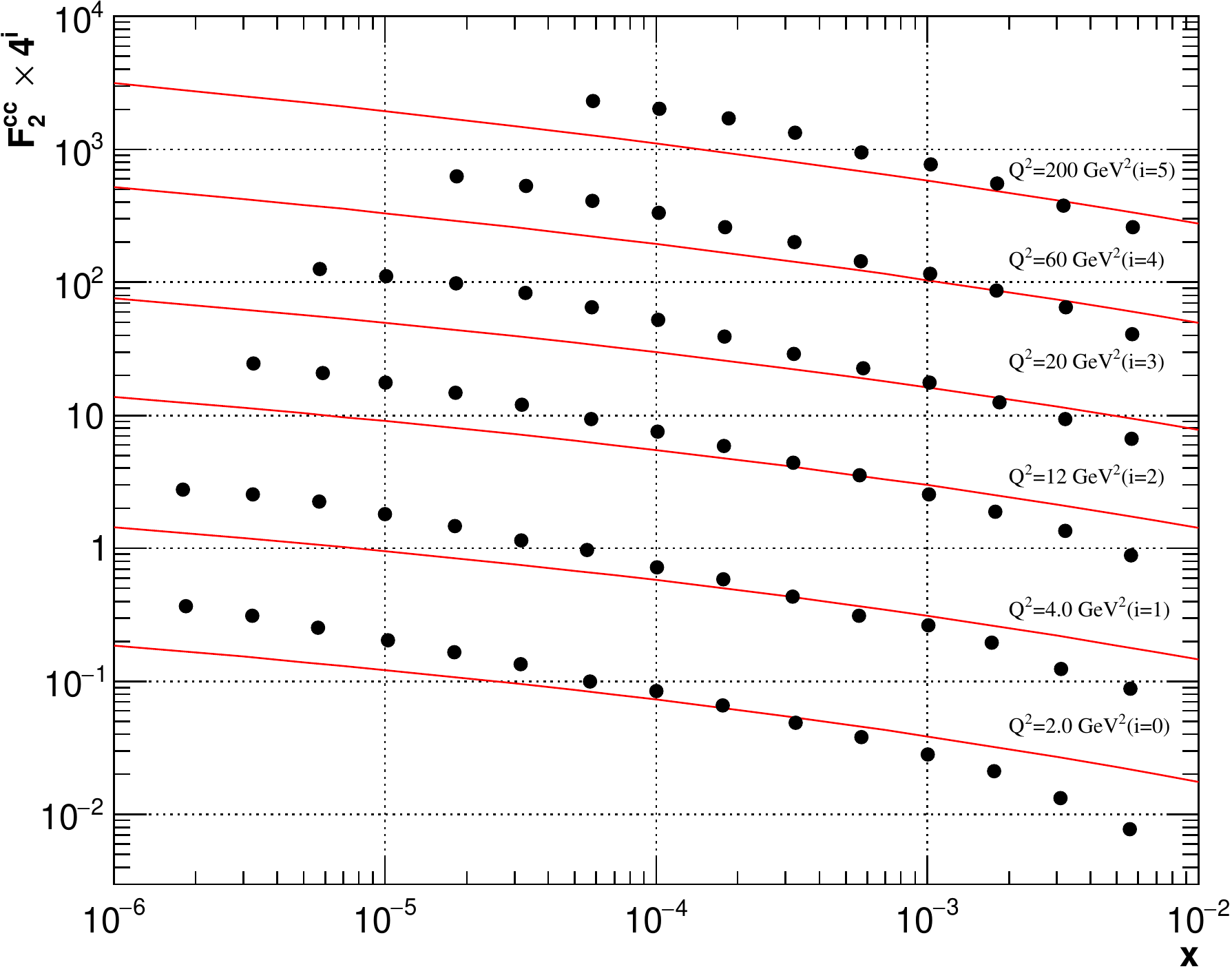}
    \includegraphics*[scale=0.45]{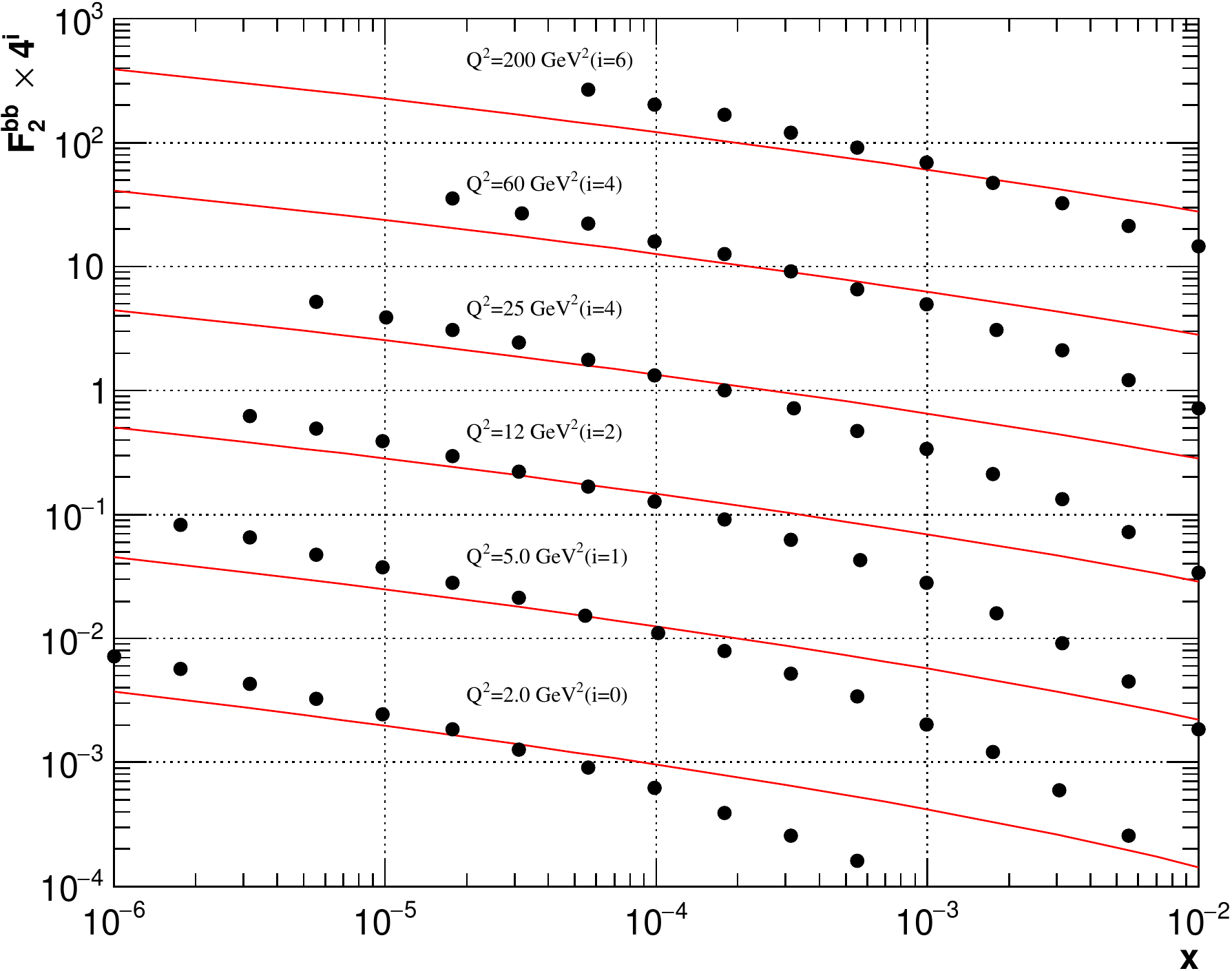}
    \caption{$F_{2}^{cc}$ and $F_{2}^{bb}$ predictions (model $V_{3}B_{2}$) at LHeC, in the range $10^{-6}\leqslant x\leqslant 10^{-2}$ and 2.0 GeV$^{2}$ $\leqslant Q^{2} \leqslant 200$ GeV$^{2}$ alongside pseudodata extracted from the Monte Carlo RAPGAP (Figs. 3.23 and 3.24 of Ref. \cite{AbelleiraFernandez:2012cc}).}
    \label{fig:F2c_LHeC}
\end{figure}

As a last analysis, based on $V_{3}B_{2}$ variant, we give predictions for the Large Hadron Electron Collider (LHeC) \cite{AbelleiraFernandez:2012cc}, which extends the kinematical range of $ep$ DIS to very low-$x$. It is proposed as a configuration with electrons of $50-100$ GeV colliding with $7$ TeV protons in the LHC accelerator. It is also  planned high energy/luminosities configuration in a long term period \cite{Br_ning_2019,Bordry:2018gri} (HE-LHeC, $\sqrt{s}_{ep}\simeq 1.7$ TeV, and FCC-ep with $\sqrt{s}_{ep}\simeq 3.5$ TeV). This allows us to explore Bjorken-$x$ in DIS down to $\sim 10^{-6}$ with high luminosity. Specifically, here we consider the LHeC scenario with $E_e=50$ GeV on $E_p=7$ TeV, $\sqrt{s}_{ep}\simeq 1.3$ TeV, with a luminosity of 50 fb$^{-1}$. This provides access to a kinematic region of  $2\times 10^{-6}< x < 0.8$ and $2 < Q^2 < 10^5$ GeV$^2$.
\begin{figure}[H]
    \centering
    \includegraphics[scale=0.5]{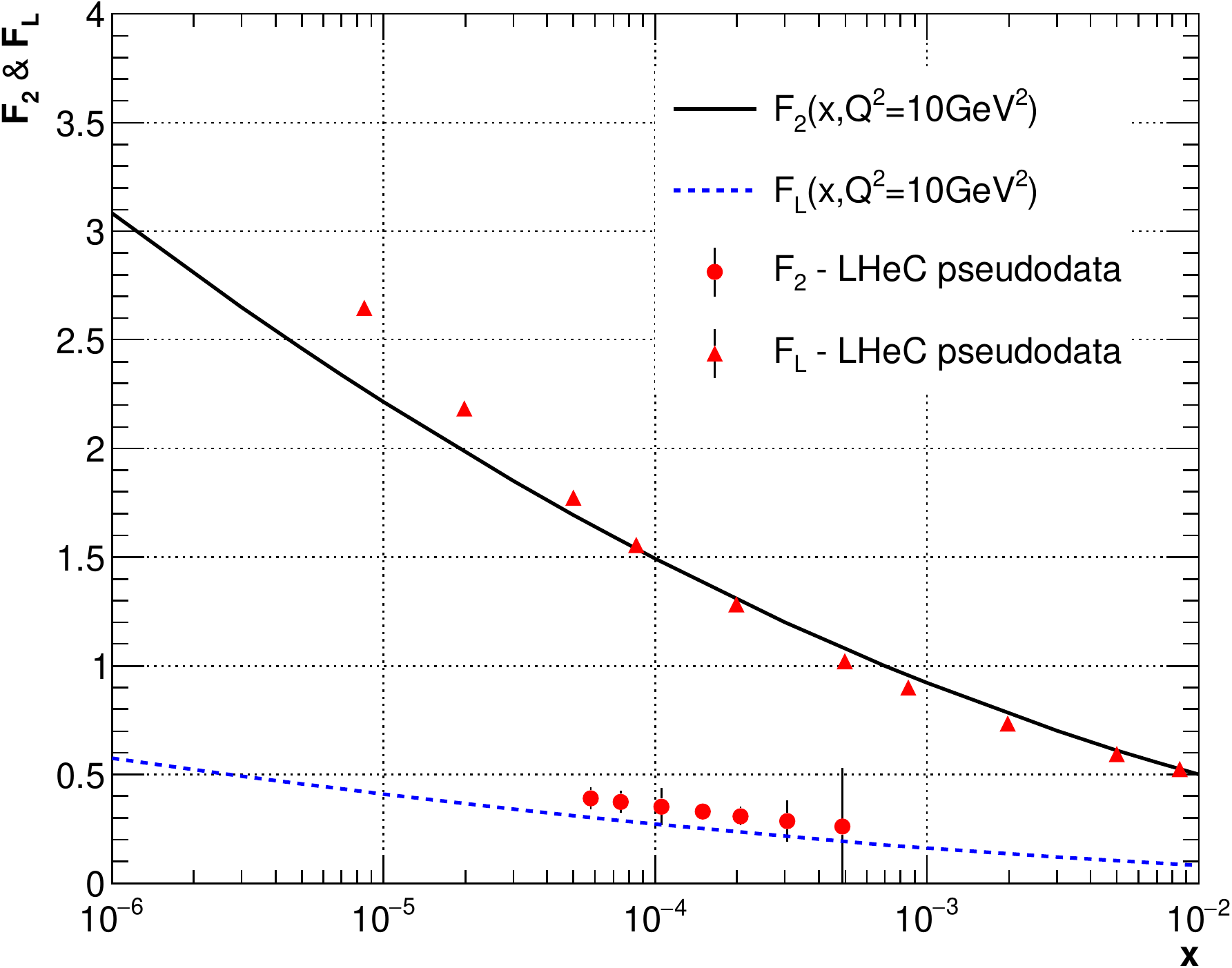}
    \caption{$F_{2}(x,Q^{2})$ and $F_{L}(x,Q^{2})$ predictions at LHeC for $Q^{2}=10$ GeV$^{2}$ and $10^{-6} \leqslant x \leqslant 10^{-2}$. Pseudodata within this kinematic window was extracted from predictions of the Monte Carlo RAPGAP in Fig. 4.13 of Ref. \cite{AbelleiraFernandez:2012cc}.}
    \label{fig:F2FL_LHeC}
\end{figure}
Our predictions to $F_{2}$ and $F_{L}$ are shown in Fig. \ref{fig:F2FL_LHeC} (fit including $c$ and $b$ quarks) for $F_2$ and $F_L$ compared to the simulated LHeC pseudodata  electron-proton collisions at $Q^2 = 10$ GeV$^2$ and for $10^{-6} \leq x \leq 10^{-2}$ \cite{AbelleiraFernandez:2012cc}. The extension of present model to very low-$x$ is reasonably consistent with simulated LHeC data and it is expected that the real measurements can be able to discriminate between the models including saturation physics and constraints on the small $x$ QCD dynamics. Predictions are also shown for the charm and bottom structure functions in Figs. \ref{fig:F2c_LHeC}. They are compared to the pseudodata generated by RAPGAP Monte Carlo for an LHeC scenario with electrons with $E_e=100$ GeV and protons with $E_p=7$ TeV   for an integrated luminosity of ${\cal{L}}_{int}=10$ fb$^{-1}$. We present the pseudodata for the configuration where the detector acceptance covers the whole polar angle range as well as events where at least one heavy quark ($Q=c,\,b$) is found with polar angles $\theta_Q > 2 \,(10)$ degrees. The overall trend of simulated data indicates a possible enhancement of the charm and bottom within the proton at very low-$x$. For the time being, bearing in mind the recent HERA results on $F_2^{c\bar{c}}$ and $F_2^{b\bar{b}}$, it seems premature (if not speculative) to take for granted such a behavior at LHeC, reason for what our predictions are shown. Future, actual data, shall shed light on this matter.

\section{Conclusions}
\label{sec:conc}
In this work we revisited and updated the AGBS color dipole model in the momentum space framework. The amplitude contains the BFKL dynamics at large $k$ (diffusion) and transition to
saturation regime using the traveling wave solutions of BK equation at leading order.
 The parameters have been fitted to the reduced cross section $\sigma_r$ \textbf{\cite{Abramowicz:2015mha}} measured at DESY-HERA, taking into account the heavy quark contributions in the theoretical prediction for proton structure function $F_2$. The investigation covered data in the region $x\leq 10^{-2}$ and $Q^2\leq 150$ GeV$^2$. An excellent quality of fit was found with $\chi^2/\mathrm{d.o.f.}\approx 1$ and good statistical significance with $p$-value either large. Using a confidence level of $95\%\, (\alpha=0.05)$, most of analyzed cases obey $p\gg \alpha$. The fit quality of the original results for the AGBS model with light (+ charm) quarks remains preserved with
 heavy flavours included. The parameters have not changed significantly in comparison with previous versions of the model, with and without heavy quarks with exception to the $\chi^{\prime\prime}(\gamma_c)$. Interestingly, the model considering only light quarks still describes the low-$x$/low-$Q^{2}$ data in a nice way. The saturation scale, $Q_s^2(x)=k_0^2x^{-v_c}$, presents a weaker growth on $x$ for heavy quarks than for only light ones. 

By using the parameters of the dipole amplitude in momentum space, ${\cal{N}}(Y,k)$, determined from the fit to the $F_2$
data, we predicted other inclusive structure functions. New predictions include the longitudinal, charm and beauty structure functions ($F_L,\,F_2^{c\bar{c}},\,F_2^{b\bar{b}}$). It is found remarkable  agreement with updated HERA data in all $Q^2$ bins. This means the model is able to emulate the DGLAP evolution at very large $Q^2$ and the correct parton saturation effects at low $Q^2$. Predictions for the LHeC kinematic range were provided and compared to available pseudodata for that TeV scale $ep$ machine.

Here, we have only considered the simplest scenario of LO  expression for dipole amplitude and an extension addressing its NLO correction could certainly be done. Recently, the first fit to HERA inclusive cross section data using the full NLO impact factor combined with an improved BK evolution has been done and the predictions are quite robust \cite{Beuf:2020dxl}. The numerical solution of the NLO BK equation presents instabilities, and  resummations of the radiative corrections are needed \cite{Ducloue:2019ezk,Ducloue:2019jmy}. These instabilities comes from subleading double logarithms arising from the incomplete cancellation between
real and virtual corrections which are Sudakov type ones. They can be resummed to all orders and  a Sudakov suppressed BK equation (SSBK)  is obtained. A fit of the reduced cross section using SSBK was shown to be reasonable \cite{Xiang:2020qtf}. Thus, it is timely to investigate the NLO evolution in a simpler phenomenological model as the AGBS one.  Moreover, we envisage as future possibility to further explore the impact parameter dependence of the amplitude at LO and NLO. The numerical solution to the BK with impact parameter dependence containing collinearly improved kernel was analyzed  in Ref. \cite{Bendova:2019psy} and  reasonable agreement with  HERA
and LHC data has been found.   Moreover, we envisage as future possibility to further explore the impact parameter dependence of the amplitude at LO and NLO. This can shed light on the $b$ dependence of the dipole amplitude in an analytical QCD model. Such study would be complementary to the numerical solution to the BK equation with impact parameter dependence containing collinearly improved kernel, which  was analyzed  in Ref. \cite{Bendova:2019psy} and  where reasonable agreement with  HERA
and LHC data has been found. Finally, the present approach can be regarded as a starting point to study diffractive DIS (DDIS) and exclusive particle production such as the deeply virtual compton scattering (DVCS) and  exclusive vector meson production, which we intend to investigate in a future work.

\begin{acknowledgments}
This work was supported by the Brazilian funding agencies CAPES and CNPq. DAF acknowledges the support of the project INCT-FNA (464898/2014-5).
\end{acknowledgments}

\bibliographystyle{h-physrev}
\bibliography{agbs_update}

\end{document}